# Role of a capping layer on the crystalline structure of Sn thin films grown at cryogenic temperatures on InSb substrates

A.-H. Chen,[1] C.P. Dempsey,[2] M. Pendharkar,[2] A. Sharma,[3] B. Zhang,[3] S. Tan,[4] L. Bellon,[5] S.M. Frolov,[3] C.J. Palmstrom,[2,*] E. Bellet-Amalric,[6] and M. Hocevar[†1]

[1)]Univ. Grenoble Alpes, Grenoble INP, CNRS, Institut Néel, 38000 Grenoble, France

[2)]Electrical and Computer Engineering, University of California, Santa Barbara, Santa Barbara, CA 93106, USA

[3)]Department of Physics and Astronomy, University of Pittsburgh, Pittsburgh, PA 15260, USA

[4)]Department of Electrical and Computer Engineering, and Petersen Institute of NanoScience and Engineering, University of Pittsburgh, PA 15260 USA

[5)]Univ Lyon, ENS de Lyon, CNRS, Laboratoire de Physique, 69342 Lyon, France

[6)]Univ. Grenoble Alpes, Grenoble INP, IRIG-CEA Pheliqs, 38000 Grenoble, France

Metal deposition with cryogenic cooling is widely utilized in the condensed matter community for producing ultra-thin epitaxial superconducting layers on semiconductors. However, a drawback arises when these films return to room temperature, as they tend to undergo dewetting. This issue is mitigated by capping the films with an amorphous layer. In this study, we examined the impact of different *in-situ* fabricated caps on the structural characteristics of Sn thin films deposited at 80 K on InSb substrates. Regardless of the type of capping, we observed that the films remained smooth upon returning to room temperature and were epitaxial on InSb in the cubic Sn ($\alpha$-Sn) phase. However, we noted a correlation between alumina capping with an electron beam evaporator and an increased presence of tetragonal Sn ($\beta$-Sn) grains. This suggests that heating from the alumina source may contribute to a partial phase transition in the Sn layer. The existence of the $\beta$-Sn phase induced superconducting behavior of the films by percolation effect. This study highlights the potential for modifying the structural properties of cryogenic Sn thin films through *in-situ* capping, paving the way for precise control in the production of superconducting Sn films for integration into quantum computing platforms.

[*]Also at California NanoSystems Institute, University of California Santa Barbara,Santa Barbara, Santa Barbara, CA 93106, USA; Also at Materials Department, University of California Santa Barbara, Santa Barbara, Santa Barbara, CA 93106, USA
[†]Electronic mail: moira.hocevar@neel.cnrs.fr

I. INTRODUCTION

a. On the cryogenic deposition of metal thin films

The development of thin films grown at cryogenic temperatures started in the 1990's. Most of the works aimed at enhancing Schottky barriers at the interface between metals and III-V's [1,2]. The idea was that by reducing the deposition temperature, interdiffusion between the metal and one of the constituents of the III-V would decrease drastically, leading to sharper interfaces. Moreover, adatom surface diffusion is also reduced, reducing surface roughness of ultra-thin films thanks to the formation of small grains. More recently, cryogenic deposition was applied to superconductor-semiconductor junctions, in the context of developing better materials and interfaces for applications in quantum information technologies and science. For lattice matched materials, epitaxial interfaces become possible, such as for Al [3]and Pb [4] on InAs, and Sn on InSb [5]. Polycrystalline films may also form for large lattice mismatch with various crystalline qualities. In general, refractory metals Nb and Ta form columnar films [6], yet Sn forms smooth polycrystalline films [7].

However, cryogenic thin films are grown far from equilibrium. They tend to dewet and form isolated islands upon returning to room temperature, similar to thin films grown at room temperature and subsequently annealed at higher temperatures [8]. One solution proposed to prevent film coalescence is to either oxidize the surface of the film or to deposit a hard amorphous cap, when still cold. For example, Al is oxidized *in-situ* [3] but it is also possible to use alumina deposited on Sn [7]. Pb cannot be oxidized because it turns completely amorphous [4]. V as well oxidizes into a bad oxide [9], and it is likely that Sn oxide has too much dielectric loss for application in quantum bits (qubits). Based on this, it is clear that the choice of the capping is contingent on the nature of the thin film and requires a systematic study.

b. On the recent interest for (cryogenic) Sn thin films

Growing focus on the superconductor-semiconductor Sn-InSb system originates from its topological properties [10,11] and potential application in quantum superconducting devices[7,12]. The α-Sn phase is predicted to exhibit multiple topological insulator and semimetal phases that appear for different film thicknesses [13–15], doping levels [16] and strain [17]. The tetragonal β-Sn phase is a superconducting metal [18,19]. Devices made of β-Sn/InSb structures exhibit properties such as hard induced gap and parity stability which are attractive for the investigation of topological superconductors, as well as for hybrid superconductor-semiconductor transmon-type devices for quantum information processing [7,20]. The superconducting gap of β-Sn is approximately 3 times larger than in Al, a popular choice for quantum devices. A larger gap comes with the possibility of stronger protection against decoherence. In this context, understanding how to control the ratio of α-to-β grains is important because mixed-phase films can lead to softening of the superconducting gap, and with it bring decoherence to qubits. What are the possible mechanisms for growing β-Sn on InSb, that is lattice-matched to α-Sn, is an interesting open question.

(Cryogenic) Sn thin films are particularly intriguing because of the allotropy of Sn. Its stable crystalline phase at room temperature is β-Sn, while α-Sn is stable at lower temperatures. The α-β transition temperature is known to be 286 K [21,22], yet some recent reports suggest a α-β transition temperature being around 307 K [23]. α-Sn grows as thin films on quasi lattice-matched substrates (InSb [14] and CdTe [13]). Those films are synthesized close to room temperature [24] and their stability originates from epitaxial matching that generates mild strain at the interface. Strain in epitaxial films being thickness dependent, the α-β transition temperature is also thickness dependent [22,25,26]. Films as thick as 200 nm transform at 343 K [26], while thinner films of 8 nm transition at 454 K [22]. In this context, the formation of superconducting thin films of β-Sn on InSb appears unexpected at nominal growth temperatures of 80 K [5,7].

c. Aim of this work

Our work addresses the question of the effect of the *in-situ* cap method on the structure of cryogenic Sn films. We found conditions that appear to correlate with the formation of β-Sn on InSb substrates. Sn films were grown at a low temperature of 80 K further favoring α-Sn. In a second step, the Sn thin films were capped either with $AlO_x$, Al or $SnO_x$. As expected, our films are predominantly α-Sn. However, when capped with evaporated alumina, they contain an admixture of β-Sn grains, based on both X-ray diffraction (XRD) and transport measurements of superconductivity. We hypothesize that this is related to the relatively higher temperature during the deposition of the AlO$_x$ cap, which is supported by an increase in the amount of β-Sn when the AlO$_x$ cap is thicker. The same capping method was applied on InSb/Sn nanowires utilized in several quantum transport measurements [7,27,28] in which superconductivity was observed systematically. Yet, the factors determining the crystal phase of Sn on nanowires may be different due to, e.g. surface roughness, impurities, restricted surface geometry and differences in heat absorption. Our enhanced understanding of cryogenic Sn thin film deposition significantly contributes to the improvement of superconducting films and hybrid materials, vital for quantum bit devices, Majorana experiments, and more. Knowing which capping technique induces phase transformation in the Sn film is crucial for favoring either the β or α phase.

II. METHODS

InSb (110) and InSb (001) substrates are heated to 638K in a chamber with base pressure below $10^{-10}$ Torr and cleaned with atomic hydrogen using a chamber pressure of $5 \times 10^{-6}$ Torr during 1h. Under those experimental conditions, X-ray photoelectron spectrometry (XPS) shows the disappearance of the InO peak as well as of the C peak[29], confirming the complete deoxidation of the substrates. The sample is then transferred *in−situ* into an ultra-high vacuum (UHV) chamber with a specially designed liquid nitrogen (LN$_2$) cooled manipulator, where the sample can be cooled to 80 K. The Sn effusion cell is heated to a temperature of 1373 K, corresponding to a deposition rate of 0.02 Å/min. Sn is grown for various times in order to study film thicknesses ranging from 5 to 50 nm.

After Sn deposition, the samples are capped using one of the three following procedures (see Figure 1a for the schematic). (1) We transfer the sampls *in vacuo* into another UHV chamber to deposit 3 nm of AlO$_x$ with an electron beam (e-beam) evaporator. (2) A 1.5 nm-thick Al film is formed using an effusion cell while still on the LN$_2$ cooled manipulator. We transfer the sample subsequently *in vacuo* to a specially designed loadlock with research grade oxygen, where it is oxidized to form a native oxide (AlO$_x$). (3) We transfer the sample into the loadlock to form a native oxide (SnO$_x$) cap.

During procedure (1), the AlO$_x$ source is heated by an electron-beam with a power ranging between 300W and 750W. We set the deposition conditions to attain a deposition rate of 6 Å/min. During procedure (2), a filament heats the effusion cell to a temperature of 1353 K, above the melting temperature of Al. The thin Al film is deposited at a rate of 6 Å/min with the chamber pressure at $10^{-10}$ Torr. During procedure (3), as well as at the end of procedure (2), the sample is transferred to the load lock kept at a pressure of 100 Torr with research grade oxygen. The oxidation of Al and Sn are self-limiting [30,31]; therefore, we expect the thickness of their respective oxides to be limited to few nm.

We analyze the crystalline structure of the Sn thin films by XRD with a Smartlab Rigaku diffractometer equipped with a Cu rotating anode. The X-ray wavelength is $CuK\alpha_1$ = 0.15406 nm. In the out-of-plane configuration, we probe the crystalline planes parallel to the substrate surface. The out-of-plane setup is equipped with a 2 bounce Ge(220) monochromator for incident beam and a long parallel slit collimator 0.228° as diffracted beam analyzer. In the grazing incidence in-plane configuration, we probe the planes normal to the substrate surface. The in-plane optical setup consists in a parallel slit collimator 0.5° for the primary and secondary optics. No monochromator was used in this configuration to maximize the intensity. The implementation of a Ni $\kappa_\beta$-filter is clearly visible in Figure 2 by the shape of the InSb peak. In the out-of-plane configuration, the probed surface is of a few $mm^2$ with a penetration depth of a few microns, while in the in-plane configuration, we are in a surface sensitive configuration with a penetration depth of a few tens of nanometers and a probed surface area of 1 cm$^2$. To complete the XRD analysis, we analyze the surface of the different samples by Atomic Force Microscopy (AFM) on 15×15 $\mu m^2$ images. The roughness of the surface is evaluated by measuring the root mean square (RMS) of each image.

We carry transport measurements to determine the thin films resistance properties at low temperature. The experiments are performed in an Oxford DR200 refrigerator at the base temperature of 20 mK. The current flowing through the sample is recorded with a 3-terminal current bias $I_{bias}$ configuration for comparative data between samples. The Resistance

Capacitance filters employed for reducing high frequency noise from exterior environment are custom made at the University of Pittsburgh. The passive second order low pass filters have resistance and capacitance of first and second stages as 820 Ohms, 1.2kOhms and 22nF, 4nF respectively. The designed 3dB cutoff frequency is about 7kHz.

III. RESULTS

a. Crystalline structure of the Sn layers grown on InSb.

Using out-of-plane XRD, we first determined the crystalline nature of 16.3 nm-thick Sn layers grown on InSb(110) and further capped using one of the three methods presented in the Methods section. The $2\theta/\theta$ XRD spectra show a main peak surrounded by fringes on Figure 1(b). The main peak corresponds to the InSb(220) reflection at $2\theta = 39.35°$. The $\alpha$-Sn peak is at a slightly lower $2\theta$ angle. Thickness fringes called Pendellesung fringes are visible around the main peak and result from the high homogeneity of the Sn film. The difference in lattice parameters between $\alpha$-Sn and InSb is 0.29% [32], therefore we expect the $\alpha$-Sn film to be strained on InSb. This was confirmed by the reciprocal space maps around both ($44\bar{4}$) and (800) diffraction peaks (see supplementary data).

The thickness of the film was extracted by fitting the oscillations with the SmartLab software (see inset Figure 1b). The AlO$_x$ capped Sn and *in–situ* oxidized Sn films are much thinner than the Sn film with an oxidized Al cap (Table I). A second and much less intense peak at $2\theta = 41.78°$ is visible, except for the native oxide capped sample. We attribute it to the SbSn(110) reflection, although it may be a reflection from Sb$_3$Sn$_4$ [33] because both have similar crystalline structures. We subsequently call this phase SbSn for simplicity. We did not observe any peak related to the presence of $\beta$-Sn between $2\theta = 30°$ and $2\theta = 155°$.

b. Correlation between the presence of $\beta$-Sn grains and the capping method

Then, we employed the in-plane XRD configuration to probe the crystalline structure of the thin films normal to the surface. The incident angle was chosen to maximize the signal from the film (and to minimize the intensity of the InSb substrate). Figures 2a-c show the $2\theta\chi-\phi$ spectra of the three 16.3 nm thick samples along the in-plane InSb[$\bar{2}$20] direction (see supplementary data for the InSb[004] direction). As the film is strained on the substrate, we did not see the α-Sn(220) reflection because it overlaps with the main InSb(220) reflection. A SbSn peak is visible, similarly to out-of-plane measurements. In contrast to the out-of-plane configuration, additional peaks is visible at lower angles thanks to the surface sensitivity of the in-plane configuration. In particular, the AlO$_x$ capped sample (Figure 2a) showed two additional peaks. The first one is attributed to the SnO(101) reflection and the second one to the β-Sn(200) reflection, respectively (Table II). The second and third samples, which are capped with Al oxidized (Figure 2b) and oxidized *in–situ* (Figure 2c), respectively, exhibit the SnO(101) reflection only, and do not show β-Sn peak.

c. Correlation between β-Sn grains properties and Sn film thickness.

Next, we determined experimentally if the β-Sn phase appeared for any Sn film thickness. For that, four different samples were prepared with nominal Sn thicknesses ranging from 6 to 40 nm and subsequently capped with 3 nm-AlO$_x$. Out-of-plane XRD was performed to evaluate the quality and the crystalline orientation of the films. The XRD spectra of this series of thin films did not differ from the reference AlO$_x$ capped sample, except for the fringe frequencies from which the effective thickness of the films was extracted (see Table I and supplementary data). We noticed that all films were fully strained (see supplementary data). In-plane XRD along the InSb($\bar{2}$20) reflection is shown on Figure 3a-d for the four samples. All the Sn films contained β-Sn grains along with α-Sn. For the thinnest films, only β-Sn(200) was observed, suggesting a preferential orientation of the β-Sn grains. The relative intensity of the β-Sn(200) reflection increased with increasing Sn thickness. For the thicker films, the β-Sn(101) orientation emerged, in addition to β-Sn(200).

The distribution of the different orientations of β-Sn with respect to the substrate orientation is visible in φ scans (Figure 4a). φ-scans reveal crystal symmetry while rotating the sample in-plane. While the thin and sharp InSb(220) reflection was 2-fold symmetric, β-Sn(200), β-Sn(101) and SnO(101) were broad, less intense, and exhibited a 4-fold symmetry. The intermixing layer SbSn(110) was 2-fold symmetric like InSb(220). In order to have a 4-fold symmetry, two potential orientations are presented in Figure 4b for the β-Sn grains. In one case, the out-of-plane [010] orientation is necessary for grains having their (101) planes oriented parallel to in-plane InSb(220). In the other case, the [001] orientation is necessary for that have their (200) planes oriented parallel to in-plane InSb(220).

Interestingly, β-Sn(001) was not visible in the out-of-plane XRD configuration. Indeed, most of the (001) reflections are forbidden according to the ICDD database. The first reflection allowed is β-Sn(004) at 2θ=151.08° where the signal is very low. We found such reflection only in one sample that contained a larger amount of β-Sn (see supplementary data).

d. Correlation between the volume of β-Sn grains and AlO$_x$ deposition time

Furthermore, we evaluated the effect of longer AlO$_x$ deposition times on the crystallinity of the films. We performed out-of-plane XRD along the InSb[001] growth direction on 15 nm-thick Sn thin films deposited on InSb (001) substrates and capped with 3 nm and 10 nm thick AlO$_x$ films, respectively (Figure 5). As the substrate orientation is different from the previous samples, the main peak in Figure 5a corresponds to the InSb(004)/α-Sn(004) reflection. There is no β phase in out-of-plane XRD scans for the 3 nm-AlO$_x$ capped Sn film, similarly to what we observed in Figure 1. In contrast, two peaks that are visible at lower 2θ values for the 10 nm-thick AlO$_x$ capped Sn film correspond to β-Sn(200) and β-Sn(101), respectively (see Table II). This confirms that a longer AlO$_x$ deposition time leads to a larger fraction of β-Sn in the Sn films.

Figure 5b summarizes the different steps of the process leading to the formation of the β phase. First, α-Sn grows epitaxial on InSb, then AlO$_x$ is deposited by e-beam. This process is responsible for the increase of the sample temperature, triggering the partial transformation of α-Sn into β-Sn.

e. Transport properties of the AlOx capped Sn films

At last, low-temperature transport measurements of 16.3 nm thick Sn films were performed to check for superconductivity. Figure 6a-c show the resistivity at 20mK of the Sn thin films plotted versus the current bias. The resistivity of both the Sn thin film oxidized *in−situ* and the Sn thin film capped with Al oxidized *in − situ* remained constant over the current bias applied to the sample (Figure 6b-c) and increased while cooling from 1K to 20mK (not shown). Those two samples did not show any superconductivity. In contrast, the resistivity of the Sn thin film capped with $AlO_x$ exhibits a characteristic drop around zero current bias, in the interval −2$\mu$A and 2$\mu$A. The drop in resistivity is a consequence of the formation of Cooper pairs, which are pairs of electrons that move through the lattice without scattering. The drop in resistivity is also surrounded by sharp peaks, typical of superconductivity [34]. We see these features at temperatures beyond 3.6K where they gradually shrink towards zero bias, consistent with the critical temperature of $\beta$-Sn and significantly beyond the critical temperature of Al [18] (see supplementary information for temperature dependent transport data). Note that the resistivity does not drop to zero, either due to the 3-terminal measurement configuration or due to the high abundance of non-superconducting $\alpha$-Sn. This latter hypothesis suggested the presence of a discontinuous $\beta$-Sn film. The sample capped with oxidized Sn in Figure 6c showed a 10-fold higher resistance likely because of its smaller thickness with respect to the other samples.

## IV. DISCUSSION

We studied $\alpha$-Sn thin films deposited on InSb substrates and capped with different methods. The objective of the study was to understand under which capping conditions $\beta$-Sn grains form in the films. In the following, we discuss the homogeneity, thickness and chemical composition of $\alpha$-Sn films. Then, we focus the discussion on the $\beta$ phase identified in the Sn films.

Our results show that the Sn films are homogeneous regardless of the capping strategy. This was confirmed by the presence of Pendellesung fringes surrounding the (220)InSb reflection in the out-of-plane XRD spectra (Figure 1) and by surface roughnesses measurements lower than 1 nm (Table I). We found a mosaicity 3 times higher in the Sn film than in the InSb substrate (see supplementary data). This is an indication that structural defects, as for example dislocations or stacking faults, are present in the film[35].

Then, the actual thickness differs from the expected one, regardless of the capping (Table I). We propose that the discrepancy is due to the partial oxidation of Sn. An oxidized phase of tin (SnO) is indeed visible in the in-plane XRD spectra (Fig 2). In the case of *in−situ* oxidized Sn, the protective SnO$_x$ cap is formed via a partial consumption of Sn. In the case of AlO$_x$ capped Sn, we propose two hypothesis: oxidation occurred by the reaction of Sn with either oxygen atoms present in the vapor during AlO$_x$ evaporation [36] or oxygen in the ambient air through a porous cap. We clearly observe in Fig 3 that the amount of SnO increases with Sn thickness, while the AlO$_x$ deposition conditions remain equal. In the case of the sample capped with Al oxidized *in-situ*, the slight oxidation of Sn confirms the second hypothesis. In all cases, oxidation of the Sn film surface gives a thinner Sn film.

Furthermore, an additional phase attributed to SbSn is found in all the samples. SbSn is likely located in contact with the substrate because of its epitaxial relationship with InSb as observed in the ϕ scans on Figure 4(a). Earlier XPS studies revealed that Sb segregates in the Sn film up to 12 nm and supersedes Sn in the lattice[24]. It is thus not surprising that the crystalline orientation of the SbSn phase is identical to that of the substrate. XPS, Auger sputter depth profile and transmission electron microscopy (TEM) plan view experiments can help identify the location of the SbSn phase more precisely. However, we found that SbSn is more abundant after several months of storage in a N$_2$ box (not shown), which suggests that the interface is not at equilibrium. We cannot explain if SbSn forms at 80K or once the samples return to room temperature. In conclusion, it is necessary to prevent the formation of SbSn at the outset to prevent further aging.

An important finding of this work is that β-Sn grains appear in the α-Sn thin films after AlO$_x$ capping. Our XRD results point out (1) that the amount of β-Sn increases for longer AlO$_x$ deposition times and (2) that the β phase appears while the α phase disappears when the temperature increases above 343 K (see supplementary data for temperature dependent XRD). We believe that absorption of heat by the thin film is responsible for the appearance of the β phase.

It is important to understand how and why heating occurs on the samples during AlO$_x$ deposition. Evaporation of (ceramic) materials is known to come together with the heating of the samples [37] because of a poor transfer of heat. In e-beam evaporation, the beam heats the source. The material vaporizes by absorbing power, emitting atoms, molecules and infrared radiations. However, electron-matter interaction also generates X-rays, electrons, and ionized particles [38], which may play a role in heating but to a lesser extent [39].

We calculated the radiation power absorbed by the sample from the source (see supplementary data) by modeling the deposition setup. However, we encountered difficulty in accurately estimating the radiating surface of $AlO_x$. A recent report by Yushkov et al. [36] considers that the entire top surface of $AlO_x$ radiates heat, though evaporation takes place from a smaller area close to the size of the e-beam. Among other parameters difficult to estimate were the effective size of the e-beam and the multiple reflections of the radiations on the walls of the deposition reactor. While we deposit $AlO_x$ at relatively low rates (6Å/min), we cannot rule out the contribution of other factors producing appreciable sample heating as the kinetic and condensation energies of $AlO_x$ or the presence of ions [38,40]. Whether the factors described are sufficient to reach the phase transition temperature needs to be clarified. A systematic study of the proportion of β-Sn in the film with varying power of the e-beam source will certainly help identifying the origin of heating and controlling the rate of formation of β-Sn.

The mechanism of the α to β transition is known to be a martensitic transformation, diffusionless and thermally activated [22,41,42]. It occurs via the distortion of the cubic lattice at the position of crystalline defects into the tetragonal lattice. Moreover, martensitic transformation leads always to polycrystalline samples but not in the sense of being random [43]. The β-Sn grains indentified by XRD have preferential orientations with respect to the α-Sn orientation and the films feature structural defects. Both findings tend to support the martensitic nature of the Sn phase transition. The transformation is partial and we believe that this is due to the film being clamped onto the substrate, which hinders immediate and complete transformation of the α phase into β-Sn. An increase in the Sn film thickness reduces both the formation energy of defects [42] and the stability of the α-Sn film [44]. This explains why the amount of β-Sn increases with thicker Sn films when all other parameters are equal (Figure 3). It would be very interesting to know the spatial distribution of the β-Sn grains within the Sn film. This could not have been studied by cross section TEM due to technical limitations (see supplementary information). However, we believe plan-view TEM on lamellae prepared by mechanical and chemical processes may reveal the spatial distribution of the β-Sn grains.

Lastly, our interest is in growing superconducting β-Sn on InSb for electronic quantum devices containing gate tunable Josephson junctions. Epitaxial β-Sn cannot form directly on InSb since their crystal lattices are too distinct. The results presented in this work demonstrate our ability to produce superconducting thin films of Sn on InSb through a phase transformation of epitaxial α-Sn into β-Sn. This transformation occurs in-situ through the heating of the sample during the deposition of a robust $AlO_x$ cap, which effectively prevents dewetting of the Sn thin film. We note that InSb nanowires covered by Sn thin films and further capped by $AlO_x$ consist in 90% of β grains [7]. The same α to β transition may occur on InSb nanowires, probably at a faster rate on nanowires due to their poorer thermal conductivity [45]. Yet, we cannot exclude that the growth mechanism of Sn on the InSb nanowires facets is different.

V. CONCLUSION

This work demonstrates the role of different *in-situ* capping methods on the structural properties of epitaxial α-Sn films grown on InSb substrates at cryogenic temperature. Importantly, both the crystalline structure and the morphology of the films remain unaltered, whether by the deposition of Al and subsequent oxidation or after the oxidation of the Sn surface. In contrast, the e-beam evaporated $AlO_x$ cap induces structural transformations of the Sn films while maintaining a smooth surface. As the deposition time for $AlO_x$ increases, larger volumes of β-Sn grains form in the α-Sn films, which suggests that heating of the film occurs during $AlO_x$ deposition. This level of control is paramount, as the phase transformation occurs during capping, effectively preserving the thin film's initial morphology. Simply heating the sample can lead to other detrimental effects, such as film dewetting. Finally, the β-Sn grains/α-Sn thin film structures exhibit superconductivity. With a critical temperature of 3.6 K, our Sn thin films offer increased flexibility in the design and operation of quantum devices Majorana experiments, and more, when compared to the widely used standard aluminum.

## VI. ACKNOWLEDGMENTS

The work in Grenoble was supported by ANR HYBRID (ANR-17-PIRE-0001), IRP HYNATOQ and the Transatlantic Research Partnership. Work at the University of Pittsburgh and UCSB was supported by the National Science Foundation (NSF) PIRE-1743717.

## VII. DATA AVAILABILITY STATEMENT

The data that support the findings of this study are available within the article, its supplementary material and are openly available in Zenodo at http://doi.org/10.5281/zenodo.7581136.

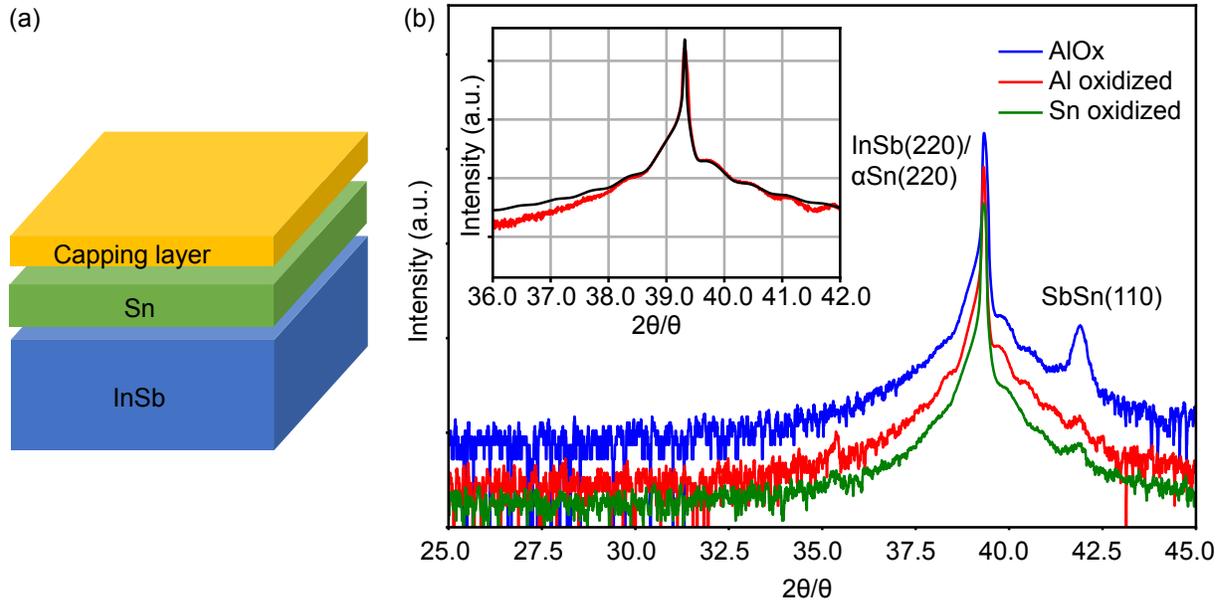

FIG. 1. <u>Determination of the crystalline phase of the Sn thin films</u>. (a) Schematic of the sample stack. (b) Out-of-plane diffraction measurement (2θ/θ scan along the InSb[110] growth direction) on the 16.3 nm-thick Sn samples grown on InSb(110) and capped with 3 nm AlO$_x$ (blue), oxidized Al and native oxide (green). Inset: magnified-view of the fitted InSb(220) peak corresponding to the oxidized Al cap sample.

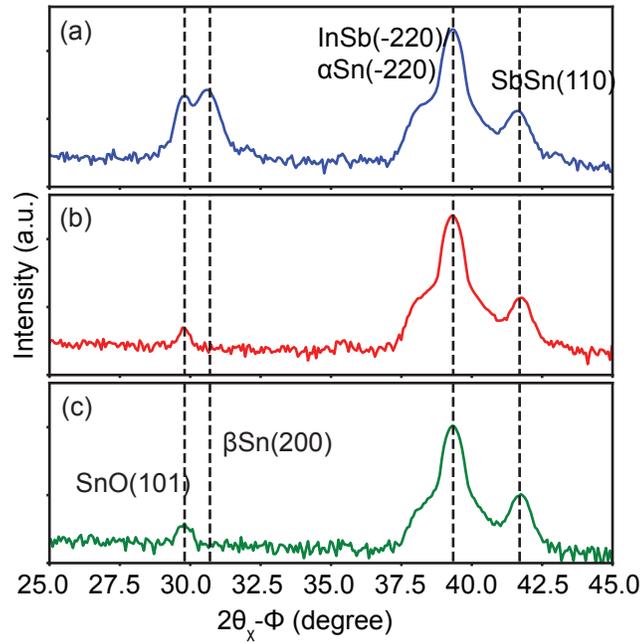

FIG. 2. <u>Influence of the capping method on the Sn films structural properties.</u> In-plane diffraction measurement ($2\theta\chi-\phi$ scans along the $[\bar{2}20]$ direction) of a 16.3-nm thick Sn thin films grown on InSb(110) and capped with (a) 3nm $AlO_x$, (b) oxidized Al and (c) native oxide. The β-Sn(200) reflexion appears only in the $AlO_x$ capped sample.

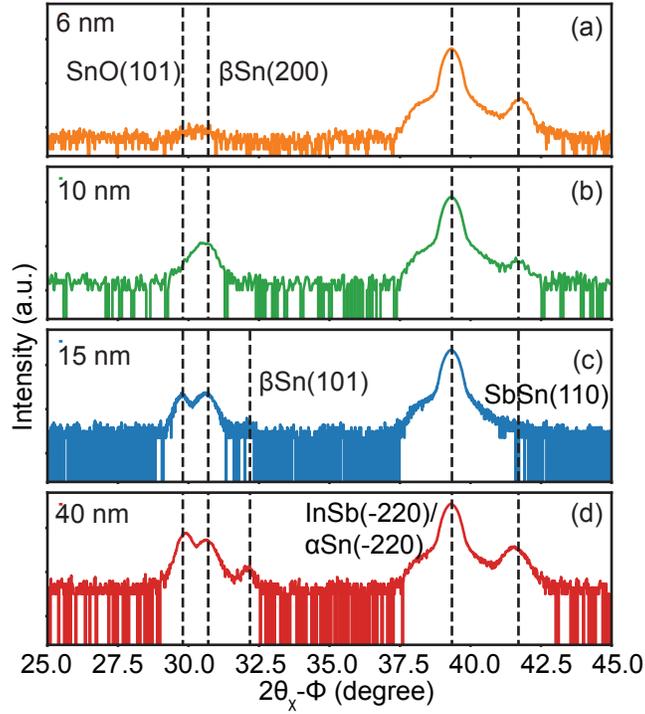

FIG. 3. **Influence of the deposited thickness on the Sn films structural properties.** In-plane diffraction measurement ($2\theta\chi-\phi$ scans along the $[\bar{2}20]$ direction) of Sn thin films of (a) 6 nm, (b) 10 nm, (c) 15 nm and (d) 40 nm thicknesses, grown on InSb(110) and capped with 3 nm AlO$_x$. The intensity of the reflections from the β-Sn and SnO phases increases with thicker Sn films.

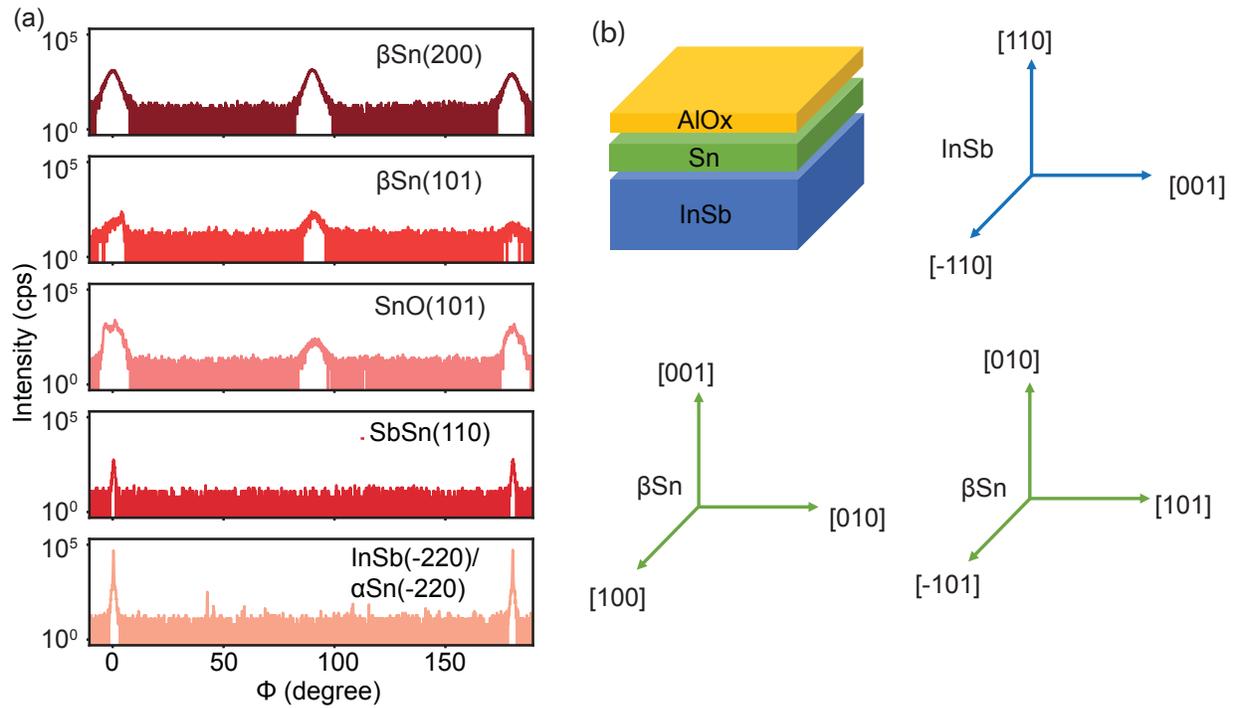

FIG. 4. <u>Orientation of the β-Sn grains in a 40nm thick Sn film</u>. (a) φ scans around the Sn(101), SnO(101), SbSn(110) and InSb(220) reflections of the 40-nm thick Sn thin film grown on InSb(110) and capped with 3 nm AlO$_x$. (b) Top left: schematic of the materials stack. Top right: crystalline orientation of the InSb(110) substrate. Bottom left and right: β-Sn(200) and β-Sn(101) grains orientation with respect to InSb(110) deduced from (a).

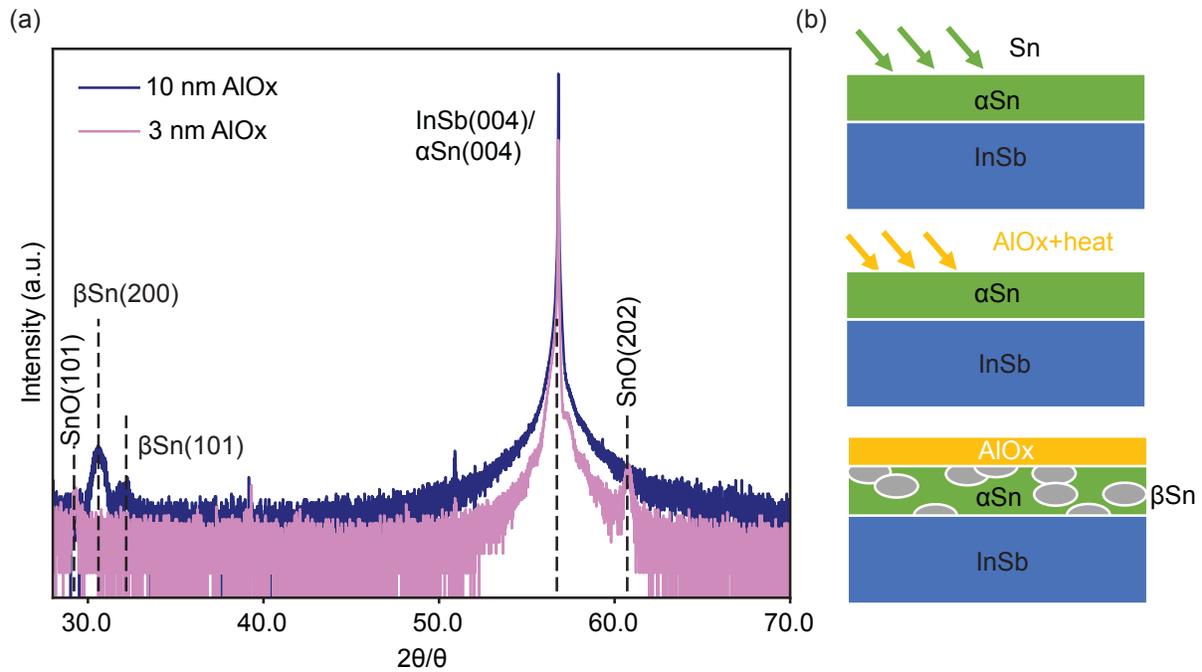

FIG. 5. <u>Influence of the AlO$_x$ thickness on the Sn films structural properties</u>. (a) Out-of-plane diffraction measurement (2θ/θ scans along the InSb[001] growth direction) of a 15 nm-thick Sn film deposited on InSb (001) substrates with 3 nm-thick AlO$_x$ cap (pink) and 10 nm-thick AlO$_x$ cap (blue). (b) Mechanism of the formation of β-Sn during AlO$_x$ deposition.

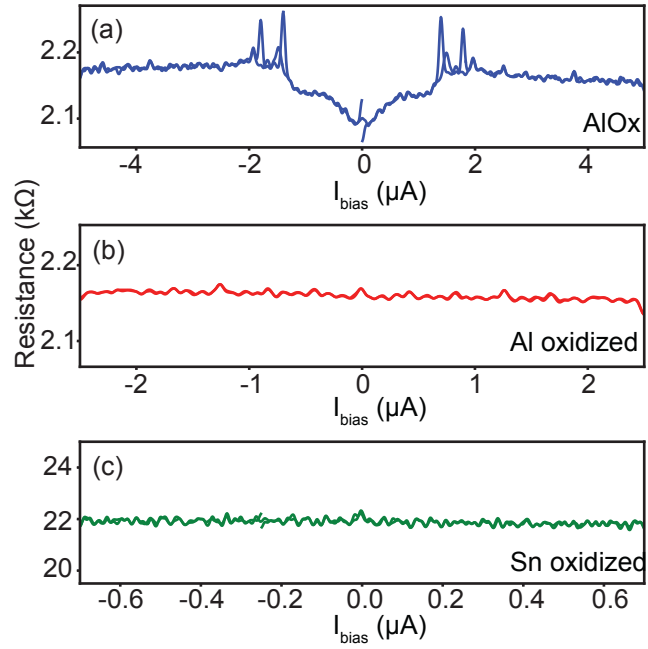

FIG. 6. <u>Superconductivity dependence on the capping method</u>. Low-temperature transport measurement at 20mK of 16.3 nm thick Sn thin films grown on InSb(110) capped with (a) AlO$_x$, (b) oxidized Al and (c) native oxide.

TABLE I. List of samples. Nominal thickness, extracted thickness from the fit of the InSb reflection in the out-of-plane configuration and surface RMS extracted from 15$\mu m$×15$\mu m$ AFM images (see supplementary data).

| Sample | Capping | Nominal thickness (nm) | Extracted thickness (nm) | Roughness (nm) |
|---|---|---|---|---|
| CPD166 (110) | 3 nm AlOx capped | 16.3 | 13.0 ± 0.5 | 0.45 ± 0.088 |
| CPD167 (110) | Al oxidized | 16.3 | 14.6 ± 0.5 | 0.63 ± 0.18 |
| CPD168 (110) | Native oxide | 16.3 | 11.3 ± 0.5 | 0.82 ± 0.19 |
| CPD125 (110) | 3 nm AlOx capped | 6 | 7.2 ± 0.4 | 0.35 ± 0.11 |
| CPD124 (110) | 3 nm AlOx capped | 10 | 10.2 ± 0.6 | 0.60 ± 0.07 |
| CPD123 (110) | 3 nm AlOx capped | 15 | 14.2 ± 0.3 | 0.65 ± 0.11 |
| CPD122 (110) | 3 nm AlOx capped | 40 | 38.1 ± 0.4 | 1.28 ± 0.20 |
| CPD123 (001) | 3 nm AlOx capped | 15 | 10.6 ± 0.3 | 0.58 ± 0.16 |
| CPD153 (001) | 10 nm AlOx capped | 15 | 12.9 ± 0.5 | 0.42 ± 0.02 |

TABLE II. List of reflections observed by in-plane XRD: measured position, ICDD value, presence in the sample

| Reflection | 2θ (°) | ICDD (°) | Presence |
|---|---|---|---|
| β-Sn(200) | 30.57 | 30.78 | all $AlO_x$ caps |
| β-Sn(101) | 32.04 | 32.18 | CPD122, CPD153 |
| SnO(101) | 29.82 | 29.90 | all except for CPD124 and CPD125 |
| SbSn(110) | 41.70 | 41.78 | all |
| InSb(220) | 39.32 | 39.35 | all |

# Supplementary information

# Role of a capping layer on the crystalline structure of Sn thin films grown at cryogenic temperatures on InSb substrates

A.-H. Chen,[1] C.P. Dempsey,[2] M. Pendharkar,[2] A. Sharma,[3] B. Zhang,[3] S. Tan,[4] L. Bellon,[5] S.M. Frolov,[3] C.J. Palmstrom,[2] E. Bellet-Amalric,[6] and M. Hocevar[1]

# Contents

# 1  List of samples studied throughout the project

TABLE I: Summary of samples.

| Sample name | Substrate orientation | Thickness Sn | Capping | XRD* | b-Sn grains | Superconducts | TEM |
|---|---|---|---|---|---|---|---|
| MP682 | (001) | 15 nm | 3 nm AlO$_X$ | In-plane | Yes | | |
| MP683 | (110) | 15 nm | 3 nm AlO$_X$ | In-plane | Yes | | |
| CPD122 | (001) | 40 nm | 3 nm AlO$_X$ | In-plane | Yes | | yes |
|  | (110) | 40 nm | 3 nm AlO$_X$ | In-plane | Yes | | |
| CPD123 | (001) | 15 nm | 3 nm AlO$_X$ | In-plane | Yes | Yes | |
|  | (110) | 15 nm | 3 nm AlO$_X$ | In-plane | Yes | | |
| CPD124 | (001) | 10 nm | 3 nm AlO$_X$ | In-plane | Yes | | |
|  | (110) | 10 nm | 3 nm AlO$_X$ | In-plane | Yes | | |
| CPD125 | (001) | 6 nm | 3 nm AlO$_X$ | In-plane | Yes | Yes | |
|  | (110) | 6 nm | 3 nm AlO$_X$ | In-plane | Yes | | |
| CPD153 | (001) | 15 nm | 10 nm AlO$_X$ | out-of-plane | Yes | | |
| CPD155 | (001) | 15.1 nm | 10 nm AlO$_X$ | out-of-plane | Yes | | |
| CPD156 | (110) | 40 nm | 10 nm AlO$_X$ | out-of-plane | Yes | | |
| CPD166 | (001) | 16.3 nm | 3 nm AlO$_X$ | In-plane | Yes | | yes |
|  | (110) | 16.3 nm | 3 nm AlO$_X$ | In-plane | Yes | Yes | |
| CPD167 | (001) | 16.3 nm | Oxidized 1.5 nm Al | In-plane | No | | |
|  | (110) | 16.3 nm | Oxidized 1.5 nm Al | In-plane | No | No | |
| CPD168 | (001) | 16.3 nm | Oxidized Sn | In-plane | No | | |
|  | (110) | 16.3 nm | Oxidized Sn | In-plane | No | No | |

* all samples were measured out-of-plane. CPD153 to CPD156 were measured only out-of-plane

## 2 TEM observation of a Sn/InSb lamellae prepared by cryo-FIB

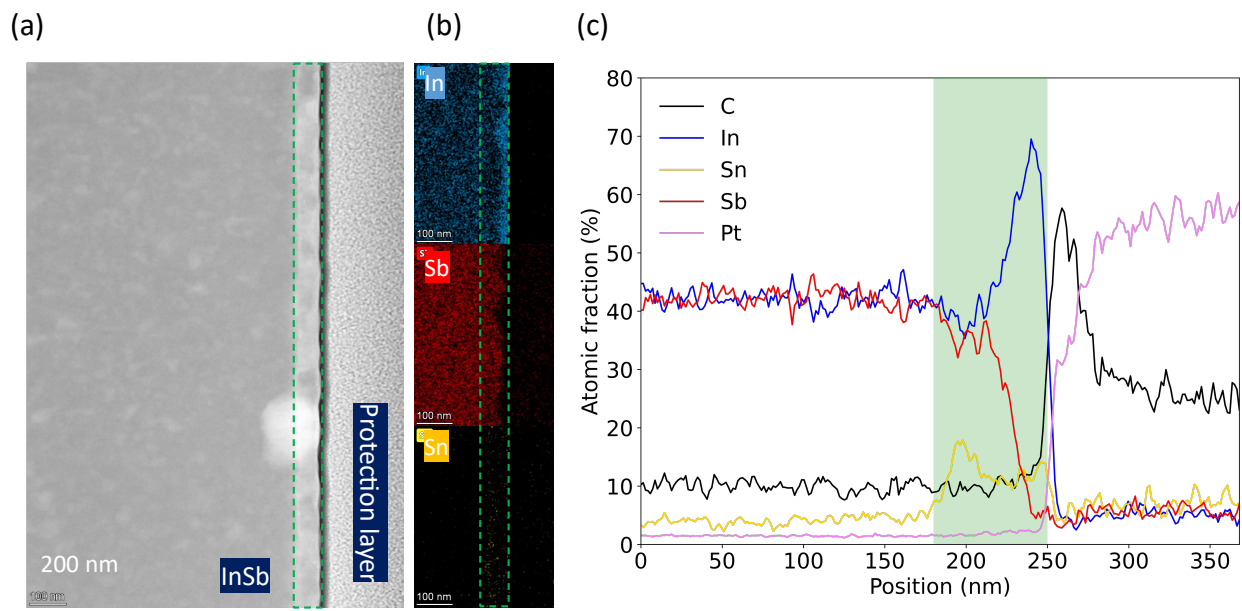

FIG. S1: [3nmAlO$_X$/40nm Sn/InSb (110)] (a) Scanning transmission electron micrograph (STEM) using an high-angle annular dark field detector shows the cross-section of the sample. From left to right: InSb (110) substrate, Sn thin film marked in green dashed lines and the protective carbon layer. (b) Electron-dispersive X-ray spectra (EDS) maps from image (a). (c) Evolution of the atomic fraction of each element along the cross-section. The blue region corresponds to the position of the Sn thin film.

We prepared a TEM lamella using a cryo-focused ion beam milling at 83K to prevent Sn melting and sputtering away by heat during the milling process. The EDS maps reveal intermixing between Sn and InSb. FigureS1 shows the EDS-TEM analysis on the Sn film grown on InSb (110) substrate. However, we find that indium migrates towards the sample surface and Sb segregates in the Sn thin film. Pure Sn cannot be detected on this lamella, which is incoherent with the results of XRD measurements. We suspect that the reason of intermixing is due to lamella preparation or the ageing of sample. Our technique does not prevent migration of elements despite the low temperature took place.

# 3 AFM observation of the sample surface

On Figure S2, we see the AFM images of Sn thin films grown on either InSb (110) or InSb (001), capped with different methods. We see grains on the InSb (110) samples. We associate them to the formation of In droplets during atomic hydrogen deoxydation process prior to Sn growth, the deoxydation recipe being optimized for InSb (001) surfaces. The RMS presented in Table II corresponds to the surface with the grains removed by a mask in the Gwyddion software.

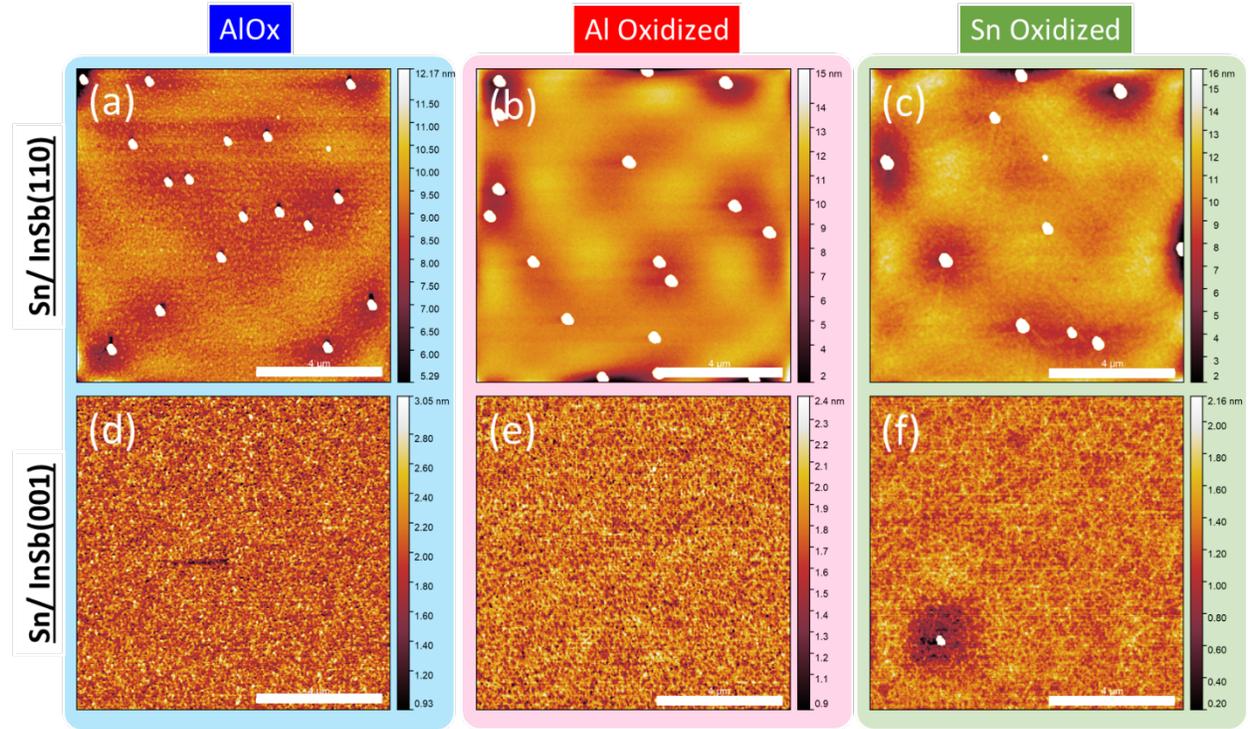

FIG. S2: [Capping layer/16nm Sn/InSb] First row: Sn/InSb (110), second row: Sn/InSb (001). From left to right: (a)-(d) AlO$_X$ capping, (b)-(e) Al oxidized capping and (c)-(f) Sn oxidized capping. The scale bar is 4$\mu$m.

TABLE II: Summary of RMS from AFM measurement.

|  | Capping | Substrate | RMS entire image (nm) | RMS excluding grains (nm) |
|---|---|---|---|---|
| CPD166 | AlO$_X$ | InSb (110) | 2.92±0.10 | 0.74±0.048 |
| CPD167 | Al oxidized | InSb (110) | 5.15±0.27 | 1.16±0.039 |
| CPD168 | Sn oxidized | InSb (110) | 6.26±0.12 | 1.21±0.28 |
| CPD166 | AlO$_X$ | InSb (001) | 0.64±0.29 | 0.75±0.44 |
| CPD167 | Al oxidized | InSb (001) | 3.17±1.88 | 0.96±0.47 |
| CPD168 | Sn oxidized | InSb (001) | 5.17±3.25 | 1.48±1.13 |

# 4 Additional XRD analysis

## 4.1 Presence of β-Sn in Sn/InSb (001) samples

We are interested in the crystalline phase of Sn thin films with various film thicknesses (CP122 to CPD125) grown on InSb (001) substrates. FigureS3 shows the in-plane XRD measurements along the InSb (220) direction of Sn films with different thicknesses grown on InSb (001) substrates. The most intense peak at $2\theta = 39.35°$ is attributed to InSb (220)/α-Sn (220) diffraction. We observe that the diffraction β-Sn (200) emerges in all the thicknesses of Sn film, albeit β-Sn (220) diffraction is observed only in thickest Sn film. SbSn (110) is observed and indicates the intermixing of Sn films and InSb substrates. We see SnO(101) diffraction in thickest Sn film grown on InSb (001). In fine, β-Sn is in both Sn/InSb (110) and Sn/InSb (001) samples.

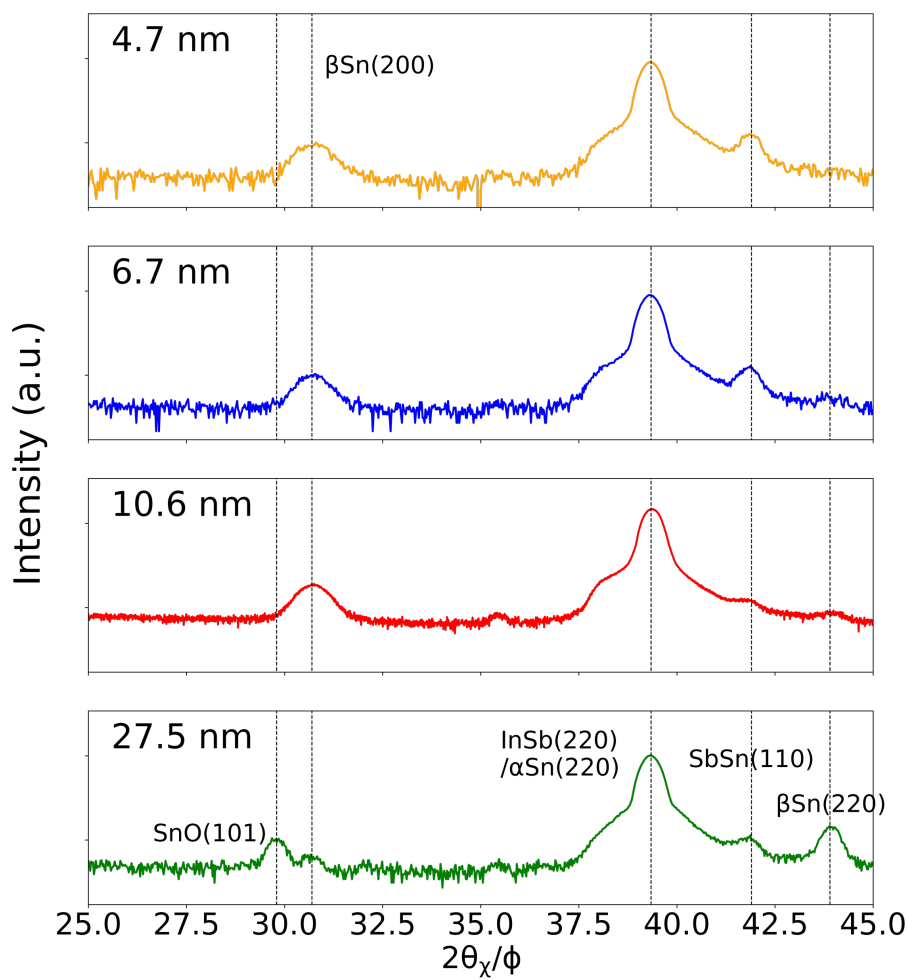

FIG. S3: [3nmAlO$_X$/Sn/InSb (001)] In-plane XRD scans on InSb (220) with different Sn thickness grown on InSb (100) substrates.

## 4.2 Presence of β-Sn (004) diffraction in out-of-plane XRD scan

Figure S4 shows the out-of-plane XRD measurement of 40nm-think Sn/InSb capped with 10nm-thick AlO$_X$ (CPD156 (110) InSb). The two intensive fringed peaks are attributed to InSb (220)/α-Sn (220) at 2θ=39.35° and InSb (440)/α-Sn (440) at 2θ=84.62° respectively. We observe a weak peak at 2θ=151.08° which corresponds to β-Sn (004). The presence of this β-Sn peak confirms the preferential orientations of the β-Sn grains observed by in-plane XRD in other samples.

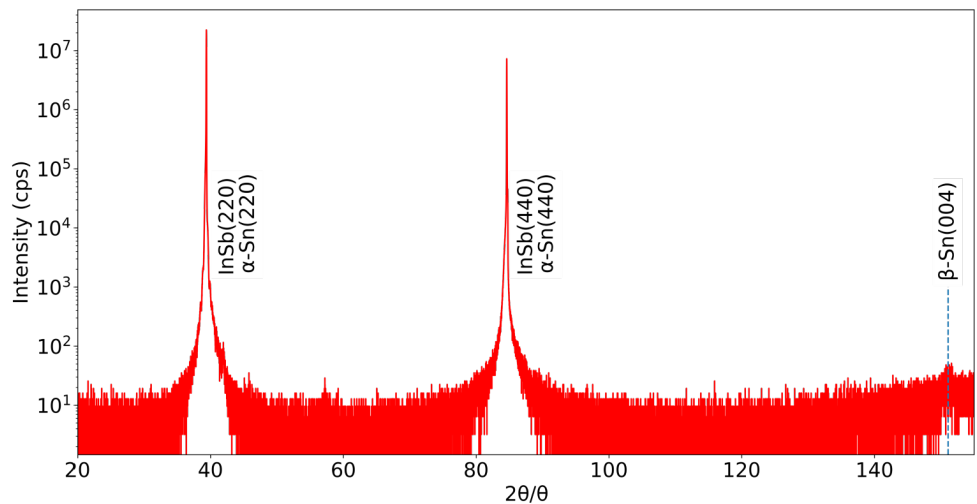

FIG. S4: [10nmAlO$_X$/40nmSn/InSb (110)] Out-of-plane XRD spectra showing the crystalline planes along the growth direction.

## 4.3 Sn/InSb(110) in-plane XRD along InSb (004)

Figure S5 shows the in-plane XRD analysis along InSb (004) rotated by 90° from the InSb (220) direction. The AlO$_X$ capped sample shows 2 peaks belonging to β-Sn grains, β-Sn (200) at 2θ = 30.78° and βSn (101) at 2θ = 32.18°. β-Sn presents a 2$^{nd}$ orientation in the α-Sn thin film along InSb (004). In both the Al oxidized capped and Sn oxidized capped samples, there are no β-Sn peaks along InSb ($\bar{2}$20) nor InSb (004). In conclusion, no β-Sn grains can be seen by XRD in the oxidized Al capped and oxidized Sn films.

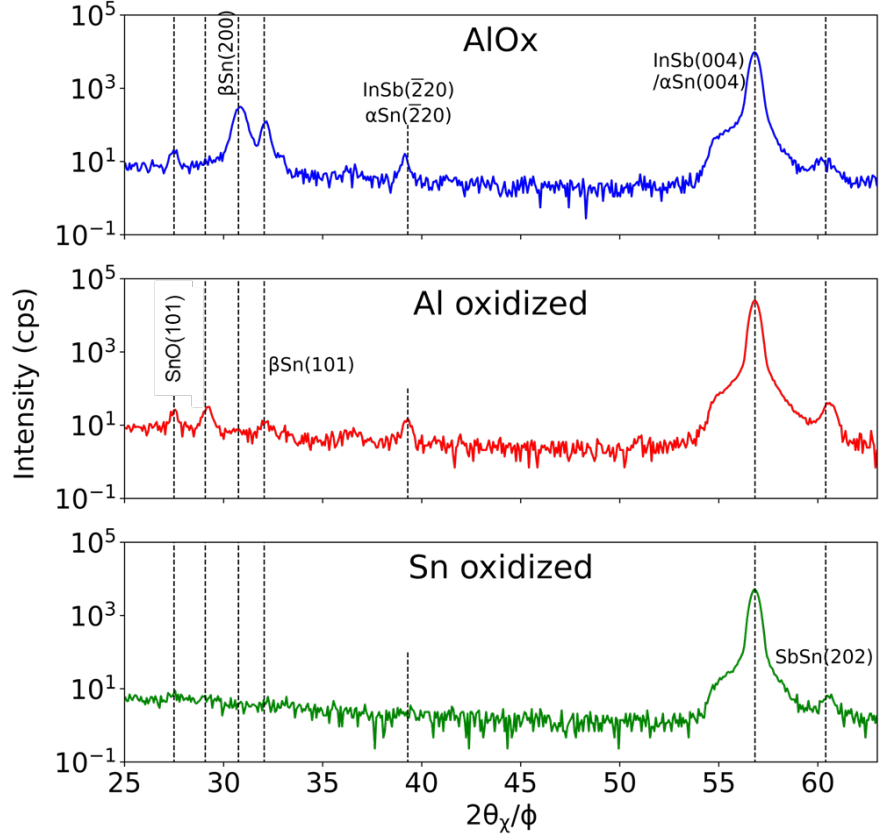

FIG. S5: [Capping layer/Sn/InSb (110)] In-plane XRD measurements along InSb (004) direction with same in-plane optics configuration for three different cappings.

We also performed the same in-plane measurements along InSb (004) on different thickness of Sn samples on InSb (110). In Figure S6, we see the substrate related reflection InSb (004) at $2\theta = 56.83°$, the β-Sn [100] reflections at $2\theta = 30.78°$ and $64.13°$, and the β-Sn [101] ones at $2\theta = 32.18°$ and $67.32°$. There is no β-Sn reflection in the 6nm-thick Sn films. However, starting from 10nm thickness, we can see the β-Sn (200) reflection. The β-Sn (101) reflection appears when the Sn thickness increases further. The spectra reveal two favorable orientations of β-Sn on InSb, (110), similar to as the in-plane spectra along InSb ($\bar{2}20$).

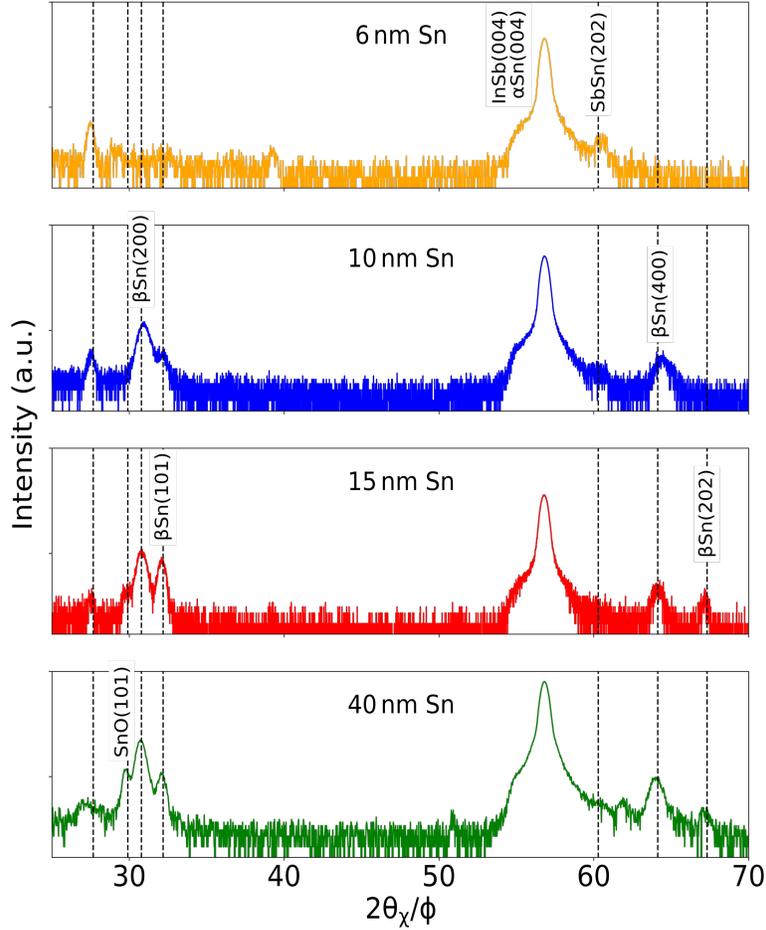

FIG. S6: [3nmAlO$_X$/Sn/(110) InSb] In-plane XRD measurements of different Sn thickness along InSb (004) direction.

### 4.4 Out-of-plane XRD on Sn/InSb(110) with various Sn thicknesses

On Figure S7, the main InSb(220) reflection features fringes. The presence of periodic fringes highlights that the α-Sn thin films on InSb (110) substrates are homogeneous, smooth and have a clean interface with InSb. Thicker α-Sn thin film exhibit fringes with shorter periods. Both α-Sn(220) and InSb(220) reflections appear at the same 2θ=39.35° due to their similar lattice parameters.

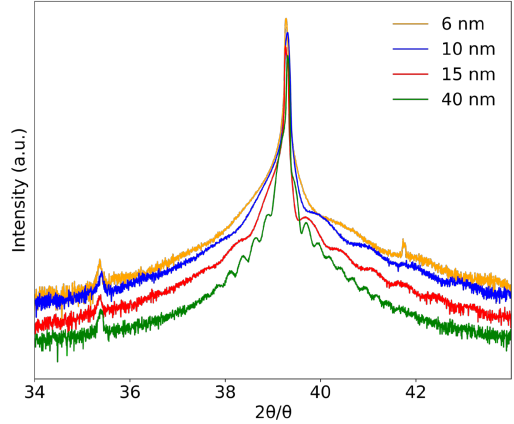

FIG. S7: Out-of-plane diffraction measurement (2θ/θ scan along the InSb[110] growth direction) of Sn samples grown on InSb(110) and capped with 3 nm. The Sn thin films are 6 nm (yellow), 10 nm (blue), 15 nm (red) and 40 nm (green) thick, respectively.

### 4.5   RSM of 40 nm Sn on InSb (110)

In Figure S8, we see the reciprocal space map of sample CPD122 with 40 nm Sn grown on InSb (110). The reflection of α-Sn ($\bar{4}44$) overlaps totally with the InSb ($\bar{4}44$) reflection in the Qx direction. It means that α-Sn is strained fully on InSb in the in-plane direction. In the Qz direction, there is an intense sharp peak from the InSb ($\bar{4}44$) reflection and a broader but weaker peak at lower Qz. The latter peak belongs to α-Sn ($\bar{4}44$). Also, α-Sn shows low mosaicity and high crystal quality on InSb as demonstrated by the low deformation of the rings (black dashed line). The elongated deformation is due to the Ewald's sphere marked in purple solid horizontal line.

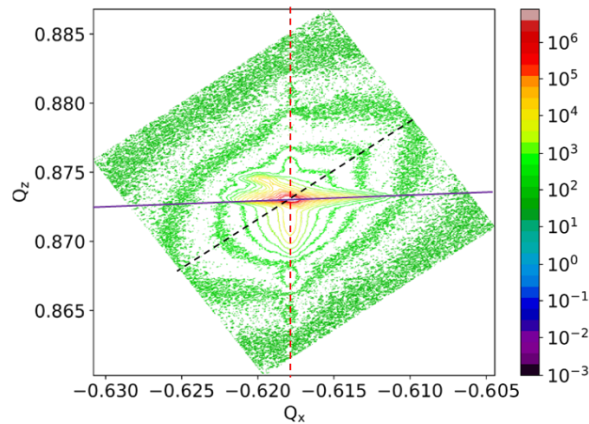

FIG. S8: Reciprocal space map at the ($\bar{4}44$) InSb reflection on 3nmAlO$_X$/40nm Sn/InSb (110)

## 4.6 Mosaicity of 40nm Sn thin film

We estimate the crystalline quality of the films by measuring the mosaicity. Figure S9 show the $\omega$ scans centered on the InSb (220) in orange and $\alpha$-Sn (220) in blue diffraction individually. We observe that the peak corresponding to the $\alpha$-Sn layer is broader than the substrate. It suggests that the crystalline quality of the $\alpha$-Sn film is not as good as that of InSb substrate. We then extracted the peak's FWHM (see Table III) from the fit by the Pseudo-Voigt equation. We find that the mosaicity of the $\alpha$-Sn thin film is three times larger than that of the InSb substrate. This increase in the FWHM of the diffraction in $\omega$-scans confirms that the crystalline quality of the thin film is lower than the InSb substrate.

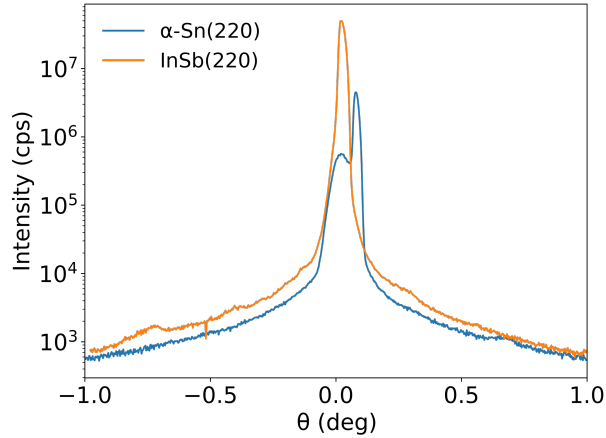

FIG. S9: [3nmAlO$_X$/40nm Sn/(110) InSb] $\omega$ scans of the (220) InSb and (220) $\alpha$-Sn reflections.

| Center of the peak | 2θ (°) | FWHM (°) |
|---|---|---|
| $\alpha$-Sn (220) | 39.27 | 0.066 |
| InSb (220) | 39.35 | 0.022 |

TABLE III: Diffraction angle 2θ and FWHM obtained from the fit of the $\omega$-scans of InSb (220) and $\alpha$-Sn (220) diffractions.

## 4.7 Reflections observed by in-plane XRD along InSb (220) and InSb (004)

TABLE IV: List of reflections observed by in-plane XRD on Sn/InSb(110): measured position, ICDD value, FWHM and relative intensity to the maximum peak of InSb (220)

| Sample | Reflection | $2\theta$ (°) | ICDD (°) | FWHM (°) | Intensity (cps) |
|---|---|---|---|---|---|
| CPD166 AlO$_X$ | $\beta$-Sn (200) | 30.58 | 30.78 | 0.90 | 712 |
| | SnO(101) | 29.79 | 29.90 | 0.41 | 476 |
| | SbSn (110) | 41.60 | 41.78 | 0.50 | 144 |
| | InSb (220) | 39.30 | 39.35 | 0.42 | 43866 |
| CPD167 Al oxidized | SnO(101) | 29.80 | 29.90 | 0.57 | 23 |
| | SbSn (110) | 41.74 | 41.78 | 0.57 | 154 |
| | InSb (220) | 39.30 | 39.35 | 0.43 | 49358 |
| CPD168 oxidized | SnO(101) | 29.79 | 29.90 | 0.52 | 12 |
| | SbSn (110) | 41.74 | 41.78 | 0.55 | 95 |
| | InSb (220) | 39.30 | 39.35 | 0.43 | 10982 |
| CPD125 6 nm Sn | $\beta$-Sn (200) | 30.31 | 30.78 | 0.80 | 6 |
| | SbSn (110) | 41.74 | 41.78 | 0.559 | 358 |
| | InSb (220) | 39.32 | 39.35 | 0.42 | 76946 |
| CPD124 10 nm Sn | $\beta$-Sn (200) | 30.57 | 30.78 | 0.653 | 499 |
| | SbSn (110) | 41.73 | 41.78 | 0.45 | 44 |
| | InSb (220) | 39.33 | 39.35 | 0.42 | 59603 |
| CPD123 15 nm Sn | $\beta$-Sn (200) | 30.59 | 30.78 | 0.57 | 615 |
| | SnO(101) | 29.78 | 29.90 | 0.40 | 508 |
| | InSb (220) | 39.32 | 39.35 | 0.42 | 58922 |
| CPD122 40 nm Sn | $\beta$-Sn (200) | 30.62 | 30.78 | 0.51 | 1677 |
| | $\beta$-Sn (101) | 32.09 | 32.18 | 0.36 | 41 |
| | SnO(101) | 29.90 | 29.90 | 0.41 | 2303 |
| | SbSn (110) | 41.55 | 41.78 | 0.67 | 504 |
| | InSb (220) | 39.32 | 39.35 | 0.43 | 49269 |

TABLE V: List of reflections observed by in-plane XRD on Sn/InSb(110): measured position, ICDD value, FWHM and relative intensity to the maximum peak of InSb (004)

| Sample | Reflection | 2θ (°) | ICDD (°) | FWHM (°) | Intensity (cps) |
|---|---|---|---|---|---|
| CPD166 AlO$_X$ | β-Sn (200) | 30.87 | 30.78 | 0.60 | 304 |
| | β-Sn (101) | 32.10 | 32.02 | 0.40 | 115 |
| | InSb (004) | 56.79 | 56.83 | 0.47 | 9683 |
| | SbSn (202) | 60.39 | 60.29 | 0.80 | 4 |
| CPD167 Al oxidized | SnO(101) | 29.22 | 29.90 | 0.60 | 31 |
| | InSb (004) | 56.86 | 56.83 | 0.45 | 25029 |
| | SbSn (202) | 60.61 | 60.29 | 0.60 | 29 |
| CPD168 oxidized | InSb (004) | 56.82 | 56.83 | 0.45 | 5269 |
| | SbSn (202) | 60.59 | 60.29 | 0.60 | 4.2 |
| CPD125 6 nm Sn | InSb (004) | 56.82 | 56.83 | 0.45 | 18896 |
| | SbSn (202) | 60.44 | 60.29 | 1.18 | 12 |
| CPD124 10 nm Sn | β-Sn (200) | 30.90 | 30.78 | 0.71 | 301 |
| | β-Sn (101) | 32.12 | 32.02 | 0.59 | 31 |
| | SbSn (202) | 60.59 | 60.29 | 0.30 | 1.2 |
| | InSb (004) | 56.82 | 56.83 | 0.46 | 26901 |
| CPD123 15 nm Sn | β-Sn (200) | 30.80 | 30.64 | 0.58 | 687 |
| | β-Sn (101) | 32.10 | 32.02 | 0.46 | 366 |
| | InSb (004) | 56.80 | 56.83 | 0.45 | 28578 |
| CPD122 40 nm Sn | β-Sn (200) | 30.72 | 30.64 | 0.56 | 1477 |
| | β-Sn (101) | 32.07 | 32.02 | 0.47 | 151 |
| | SnO(101) | 29.77 | 29.90 | 0.40 | 198 |
| | InSb (004) | 56.81 | 56.83 | 0.47 | 72190 |

## 4.8 Temperature XRD measurements on 40nm Sn thin film

FigureS10 shows temperature-dependent XRD measurements on the 40nm-thick Sn thin films grown on InSb(110). We observe that the periodic fringes disappear at around 70°C, and meanwhile the β-Sn peaks appear. This transition temperature of 40nm-thick Sn film is higher than that of bulk α-Sn (13°C). This evolution suggests that α-Sn transitions into β-Sn at this temperature. The transition from the α- to β-phase of Sn is irreversible in our temperature range.

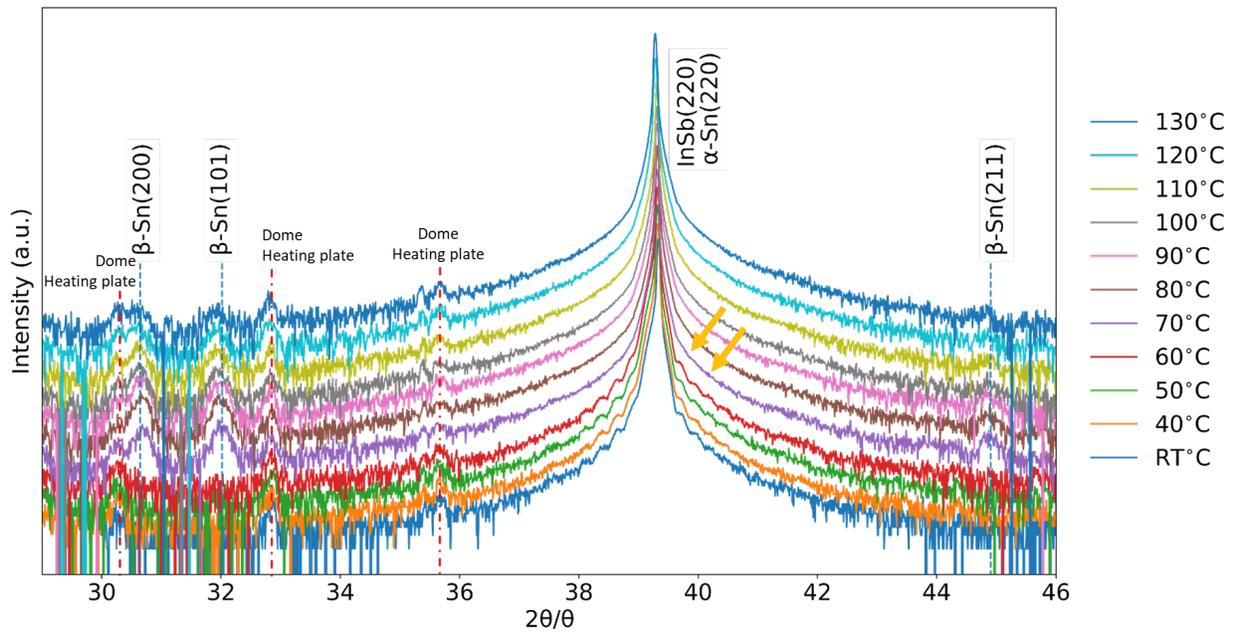

FIG. S10: Out-of-plane XRD scans performed on the 40nm-thick Sn thin film grown on InSb (110) (CPD122). The yellow arrows indicate the temperature at which the Pendellösung fringes disappear. The β-Sn reflections are visible and correspond to the blue dashed lines. The peaks originating from the setup are indicated by the red vertical lines.

# 5  Numerical estimation of the evolution of the infrared radiation of the source

To estimate the temperature evolution due to infrared radiation solely, we neglected the energy absorbed due to the conversion into heat of the condensation and the kinetic energies of the $AlO_X$ molecules (or any other energized particles such as secondary electrons, ionized molecules, etc) when they reach the sample holder surface. The heat flux by conduction from the sample holder to its support is also neglected because the holder is weakly connected (3 points of contact).

The heat balance model is based on the experimental deposition procedure. The holder on which is mounted the sample consists in a molybdenum disk with a diameter of 3inch and is 2mm-thick. The Sn/InSb sample surface has a smaller diameter 1inch and a smaller thickness of 0.5mm. The sample is glued with indium insuring excellent thermal contact. We therefore considered the holder and ignored the sample. The model includes three rates: (1) absorption of the radiation Pa emitted by the $AlO_X$ source, (2) absorption of the radiation $P_b$ emitted by the vacuum chamber at room temperature Tr ~ 300K and (3) emission of radiation Pc(T) by the sample holder itself at temperature T. The emissivity of the materials was difficult to estimate since it depends in practice on the temperature, the wavelength range, the surface treatment, etc. In our case, both the steel chamber and the molybdenum holder were oxidized, had a rough surface and were covered by materials deposited within the chamber. Therefore, their emissivity considered was in the upper bound of the emissivity range (0.5-1). For simplicity, the source, the chamber and the holder were considered to be black bodies with an emissivity of 1. Finally, the e-beam irradiated the $AlO_X$ ingots with a maximal power of 750W. The Stephan-Boltzmann law gives a temperature of ~ 2166K for an total surface of the $AlO_X$ ingot of 6x10x10mm². The heat balance of the system can be written as:

$$C(T)\frac{dT}{dt} = P_a + P_b - P_c(T), \quad (S1)$$

where C(T) is the heat capacity of the sample holder and t is the time passed since liquid nitrogen cooling stopped (the sample left the low temperature chamber in which Sn was evaporated).

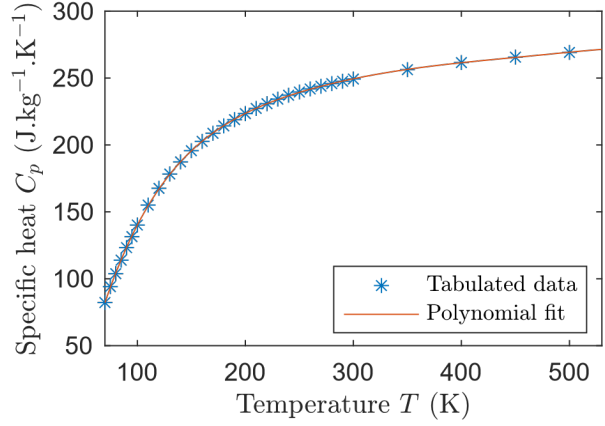

FIG. S11: Specific heat of Molybdenum: tabulated data from reference 1 and sixth order polynomial fit used to solve equation S1.

The heat capacity C(T) of the sample holder is computed from its volume (disk of radius $r_{Mo}$ = 38.1mm and thickness $t_{Mo}$ = 2mm), density of Molybdenum ($10.3 \times 10^3$ kg.m$^{-3}$), and specific heat Cp(T). The latter is estimated from a sixth order polynomial fit of the tabulated data from ref[1], as plotted in figure S11. The radiation powers Pa, $P_b$, Pc, can be estimated from the Stephan-Boltzmann law, in the black body approximation:

$$P_a = \pi r_{Mo}^2 \sigma T_{AlO_x}^4 \frac{A_{AlO_x}}{\pi d_{AlO_x}^2}, \quad (S2)$$

$$P_b = 2\pi r_{Mo}(r_{Mo} + t_{Mo})\sigma T_r^4, \quad (S3)$$

$$P_c = 2\pi r_{Mo}(r_{Mo} + t_{Mo})\sigma T^4, \quad (S4)$$

where $\sigma$ = 5.7×10$^{-8}$ Wm$^{-2}$ K$^{-4}$ is the Stephan-Boltzmann constant, $T_{AlOX}$ = 2166K is the temperature of the AlO$_X$ source at a distance $d_{AlOX}$ = 250mm from the target, and presenting an area $A_{AlOX}$ = 160$^{mm2}$ to the sample (projection on a plane perpendicular to the coating direction), and Tr = 300K is the temperature of the chamber walls. The ratio of areas in equation S2 corresponds to the solid angle of the source viewed from the target, thus to the fraction of thermal radiation that reaches directly the sample. In the black body approximation, using the direct exposure proportional to the solid angle of the source at 25cm, we obtained Pa ~ 4.6W. This is a lower bound since a fraction of the source radiation was reflected actually onto the sample by the chamber walls. From equation S3, we estimate $P_b$ = 4W. Knowing that, we used values ranging from 2 to 16W in equation S1. We perform a variable separation and an integration to solve equation S1 and get:

$$t = \int_{T_0}^{T} \frac{C(T')}{P_a + P_b - P_c(T')} dT' \tag{S5}$$

---

[1] V. Y. Bodryakov, "Correlation of temperature dependencies of thermal expansion and heat capacity of refractory metal up to the melting point: Molybdenum," High Temperature 52, 840–845 (2014).

This equation can be integrated numerically and yields the plots of Figure S12. We observe the steep rise of temperature due to the source thermal radiation, reaching the Sn-α to Sn-β phase transition temperature $T_{\alpha \to \beta}^{film}$ for a few minutes when source power and exposure time are high enough.

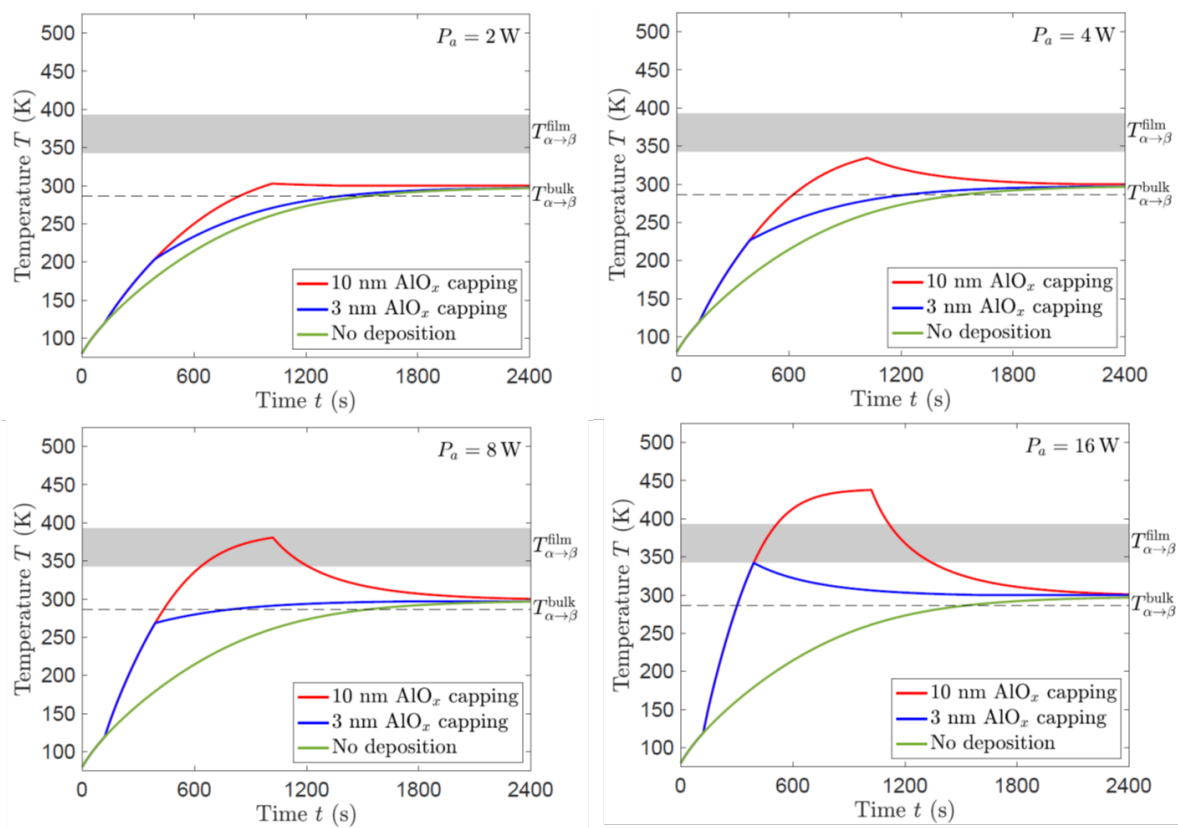

FIG. S12: Time evolution of the sample temperature during exposure to the AlOx source thermal radiation for powers Pa indicated in the upper right corner of each graph.

Without exposure to the AlO$_X$ source, the sample returned to room temperature after 30 minutes due to the radiation of the chamber at 300K. Exposure to the AlO$_X$ source radiation started at t=120s (the duration of the in-situ transfer from the low temperature chamber), and lasted 270s for a 3nm capping (900s for a 10nm capping). We observed a steep temperature rise, obviously faster for larger powers. In this qualitative picture, the sample exceeds the bulk transition temperature above which the $\beta$-Sn phase is favored for infrared radiation powers above 5W[2].

---

[2] R. F. Farrow, D. S. Robertson, G. M. Williams, A. G. Cullis, G. R. Jones, I. M. Young, and P. N. Dennis, "The growth of metastable, heteroepitaxial films of $\alpha$-Sn by metal beam epitaxy," Journal of Crystal Growth 54, 507–518 (1981).

## 6 Low-T transport measurement on Sn/InSb (001) capped with AlO$_X$

The samples were rinsed with Acetone and IPA 1min each and the copper sampled holder was wiped with IPA rinsed KIMTECH wipes before placing the samples to the holder using silver paste. The Aluminum wire bonds from gold plated copper pads on the sample holder to the samples were made using WestBond 454647E ultrasonic wedge-wedge wire bonder. All measurements are performed in DC with second generation IVVI DACs to power the current source (S4m) and voltage readout (M2b) modules at necessary/available amplification settings (10uA/V and 100V/V respectively for presented data). No lock-ins were used. Three-point measurements were performed for a fair comparison between the three samples since one of the contacts came off during cooldown. Figure S13 shows the electronic transport measurement at low temperature performed in a dilution fridge. We see that switching currents appear between $I_{bias}$ = 200$\mu$A and –200$\mu$A while the temperature decreases below 2K. Between the switching currents, we observe a relatively low-resistance region surrounding $I_{bias}$ = 0$\mu$A: it is the superconducting gap. We find that the superconducting gap closes up while the temperature increases. The non-zero resistance in the superconducting gap is attributed to the insulating a-Sn layer.

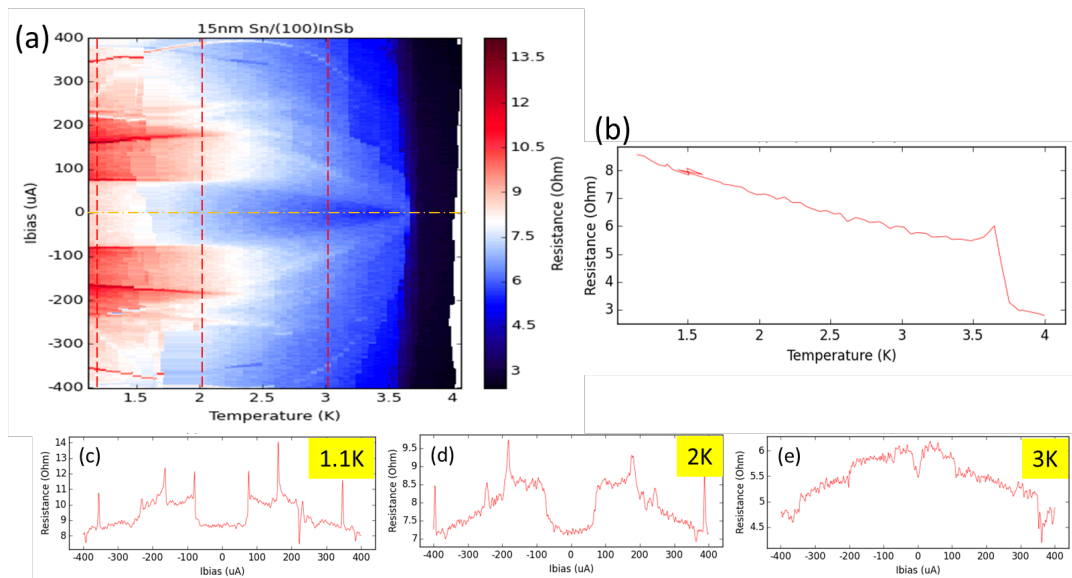

FIG. S13: [3nmAlO$_X$/15nmSn/InSb (001)] Low temperature transport measurement of the 15nm Sn thin film grown on an InSb (001) substrate. (a) 2D resistivity map as a function of $I_{bias}$ and temperature. (b) R-T line scan at $I_{bias}$ = 0$\mu$A extracted from (a) (indicated by a yellow horizontal dashed line). (c)-(e) Resistance versus $I_{bias}$ at different temperatures extracted from (a).